\documentclass[fleqn,usenatbib]{mnras}

\usepackage[T1]{fontenc}
\DeclareRobustCommand{\VAN}[3]{#2}
\let\VANthebibliography\thebibliography
\def\thebibliography{\DeclareRobustCommand{\VAN}[3]{##3}\VANthebibliography}
\usepackage{soul,color}
\usepackage{graphicx}	% Including figure file
\usepackage{amsmath}	% Advanced maths commands
\usepackage{amssymb}	% Extra maths symbols
\usepackage{newtxtext,newtxmath}
\usepackage{physics}

\usepackage{soul}
\newcommand{\Msun}{\, \mathrm{M}_{\odot}}

\title[Galactic satellite systems in CDM, WDM and SIDM]{Galactic satellite systems in CDM, WDM and SIDM}

\author[Victor J. Forouhar Moreno et al.]{
Victor J. Forouhar Moreno$^{1}$\thanks{E-mail: victor.j.forouhar@durham.ac.uk},
Alejandro Ben\'itez-Llambay$^{2}$,
Shaun Cole$^{1}$,
and Carlos Frenk$^{1}$
\\
$^{1}$Institute for Computational Cosmology, Department of Physics, Durham University, Durham DH1 3LE, UK\\
$^{2}$University of Milano-Bicocca, Piazza della Scienza, 3, 20126 Milano MI, Italy
}

\date{Accepted XXX. Received YYY; in original form ZZZ}

\pubyear{2022}

\begin{document}
\label{firstpage}
\pagerange{\pageref{firstpage}--\pageref{lastpage}}
\maketitle 

\begin{abstract}
We investigate the population of bright satellites ($M_{*} \geq 10^{5} \Msun$) of haloes of mass comparable to that of the Milky Way in cosmological simulations in which the dark matter (DM) is either cold, warm or self-interacting (CDM, WDM and SIDM respectively). The nature of the DM gives rise to differences in the abundance and structural properties of field halos.  In WDM, the main feature is a reduction in the total number of galaxies that form, reflecting a suppression of low-mass DM haloes and lower galaxy formation efficiency compared to CDM. For SIDM, the changes are structural, restricted to the central regions of haloes and dependent on the assumed self-interaction cross-section. We also consider different baryonic subgrid physics models for galaxy formation, in which supernova gas blowouts can or cannot induce the formation of a core in dwarf galaxies. Overall, the inclusion of baryons lessen the differences in the halo properties in the different DM models compared to DM-only simulations. This affects the satellite properties at infall and therefore their subsequent tidal stripping and survival rates. Nonetheless, we find slightly less concentrated satellite radial distributions as the SIDM cross-section increases. Unfortunately, we also find that the satellite populations in simulations with baryon-induced cores in CDM and WDM can mimic the results found in SIDM, making the satellite stellar mass and maximum circular velocity functions heavily degenerate on the assumed nature of the DM and the adopted subgrid modelling. These degeneracies preclude using the brightest satellites of the Milky Way to constrain the nature of DM.

\end{abstract}


%%%%%%%%%%%%%%%%%%%%%%%%%%%%%%%%%%%%%%%%%%%%%%%%%%

%%%%%%%%%%%%%%%%% BODY OF PAPER %%%%%%%%%%%%%%%%%%

\section{Introduction}

The precise nature of the dark matter (DM) is as-yet unknown, despite making up the largest fraction of the universal matter energy-density budget \citep{2020.Planck}. This is because its existence has only been inferred through astrophysical tests relying on gravitational probes, such as the rotation curves of galaxies \citep{Rubin.1970}, strong gravitational lensing \citep{Wambsganss.2004} or X-ray emission from galaxy clusters \citep{Voigt.2006}. Despite ongoing searches for a particle counterpart that could account for most of the dark matter, none have yet made a conclusive detection, directly \citep{Marrodan.2016} or indirectly \citep{Gaskins.2016}. 

Nonetheless, assuming dark matter is a heavy particle whose distribution on large scales is solely dictated by gravity results in a remarkable agreement between predictions and observations on large cosmological scales \citep{Davis.1985}. These range from the distribution of galaxies at the present-day \citep{Cole.2005,Springel.2006, Rodriguez.2016}, to the anisotropies imprinted in the Cosmic Microwave Background, back when the Universe was only 300,000 years old \citep{2020.Planck}.

A natural particle candidate satisfying these criteria are weakly-interacting massive particles (WIMPs; \citealt{Ellis.1984}). These are hypothetical particles which arise on electroweak scales -- $\mathcal{O}(\mathrm{GeV - \mathrm{TeV}})$ -- and whose predicted relic abundance is similar to the one required by the inferred DM density. Within the WIMP landscape, an exciting prospect is the lightest neutralino, a particle predicted from well-motivated minimal super-symmetric extensions to the Standard Model. These considerations make cold dark matter (CDM) the \textit{de facto} DM model. However, no direct evidence for supersymmetry \citep{Canepa.2019} or WIMP-like dark matter candidates \citep{Aprile.2018} has been detected yet. As more of the plausible parameter space is excluded, we may need to re-visit our expectations on what the particle nature of the DM is.

 Nonetheless, there are other well-motivated models which have not yet been ruled out. One such example is warm dark matter (WDM), a lighter particle than CDM with masses in the keV range. A promising WDM particle is the sterile neutrino, which is a hypothetical right-handed equivalent of the Standard Model (SM) neutrino. These arise naturally in many Grand Unified Theories (e.g. \citealt{Pati.1974}) and could provide a natural explanation for the small mass of SM neutrinos via the see-saw mechanism \citep{King.2015}. Cosmologically, its lighter nature entails its free-streaming length -- the spatial scale over which primordial density perturbations are erased -- is larger than in CDM. Consequently, its power spectrum is suppressed at small spatial scales relative to CDM. This has a number of interesting consequences, from a decrease in the number of low mass haloes to a delay in their formation time. The latter effect also results in structural changes in the distribution of dark matter haloes, such as lower concentrations. Thus, WDM is able to reproduce the success of CDM on large scales, whilst modifying the predictions on smaller scales. 

Another alternative is a particle that is able to scatter elastically with itself, self-interacting dark matter (SIDM). Although initially proposed to solve the so-called missing satellites and cusp \textit{versus} core `problems' \citep{Spergel.2000}, there are several particle physics models that naturally result in self-interactions between DM particles (e.g. \citealt{McDonald.2002,Buckley.2010}). This leads to changes in the velocity and density profiles of the central regions of haloes, turning their cuspy NFW-like distributions to cored isothermal ones. Moreover, if the cross-section is large enough, core-collapse can be triggered and revert the flat density core to a super-cusp. Although the largest velocity-independent cross-sections are likely ruled out based on cluster-mass constraints \citep{Peter.2013,Rocha.2013}, there is still the possibility of large cross-sections at low masses via velocity-dependent cross-sections. Nonetheless, it is worth noting that many of the previous constraints have been overstated to some degree, because off simplifying assumptions, limited cluster statistics, or a lack of baryons in the simulations from which the constraints are derived \citep{Robertson.2018}.

The above changes to the DM model thus primarily alter predictions on small scales, either in the abundance of low mass structure or the distribution of DM in the centre of haloes. Consequently, we need to test these models in an appropriate environment where these changes are observationally accessible. An excellent test bench for this is the Local Group. This is because surveys such as SDSS, DES and ATLAS have made possible the discovery of low-surface brightness objects that probe the edge of galaxy formation \citep{Torrealba.2016}. Moreover, GAIA offers a unique view into the kinematics of some of these objects, leading to the discovery of the `feeble-giants' Antlia II \citep{Torrealba.2019}, whose properties are difficult to explain in a Universe dominated by collisionless dark matter \citep{Caldwell.2017,Fu.2019, Borukhovetskaya.2022}.

Objects orbiting around larger, more massive ones are subject to gravitational tides, which strip dark matter from their haloes. At fixed orbital parameters, the efficiency of this process depends sensitively on the internal structure of the DM haloes \citep{Penarubia.2010}. Thus, differences in the satellite's underlying inner dark matter distribution are amplified, leading to very different satellite populations based on their survivability. Thus, this suggests that in principle, we may indirectly probe the nature of dark matter by comparing the properties of the present-day population of satellites around the Milky Way (MW) to the results of hydrodynamical simulations. 

For the purposes of this study, the inclusion of baryons is paramount for reliable predictions. Firstly, it allows a more meaningful comparison to observations, since not all DM halos host galaxies. Secondly, the processes associated with galaxy formation and evolution can alter the global properties of haloes and how dark matter is distributed within. These effects are mass dependent and could, in principle, be degenerate with changes to the DM model \citep{Khimey.2021, Burger.2022}, e.g. core formation driven by supernovae-driven gas blowouts \citep{Navarro.1996b,Read.2005} \textit{vs} self-interacting dark matter. Moreover, the presence of a disk, and subsequent contraction of the DM halo, can greatly enhance the destruction of subhaloes \citep{Sawala.2017,Garrison-Kimmel.2017, Richings.2020}. 

Limits on available computational power means we need to resort to subgrid implementations to model baryonic physics when simulating galaxy formation in a cosmological setting. Although they are able to make realistic predictions once calibrated \citep{Genel.2014, Schaller.2015, Ludlow.2017, Hopkins.2018}, there are different parametrisation choices and many of their free parameters can be degenerate with others. This can result in different predictions on yet unconstrained relations, such as the properties of the IGM \citep{Kelly.2021}. 

One such example particularly relevant to stripping is whether supernovae-driven gas blowouts are able to form cores in dwarf galaxies. Depending on the choice of subgrid parameters, simulations produce dwarfs with central density cores (FIRE, \citealt{Onorbe.2015}; NIHAO, \citealt{Tollet.2016}) or not (EAGLE and AURIGA, \citealt{Bose.2019}). The definitive or insufficient evidence for the existence of cores in dwarf galaxies is hotly debated, with some attributing their inferred presence to difficulties in the kinematic modelling \citep{Oman.2019,Roper.2022}. Nonetheless, it is important to consider both possibilities, especially from the point of view of disentangling baryonic effects from different DM models.

Given the all of the above, this paper sets out to study how the properties of the satellite systems of haloes with masses similar to our MW -- within a factor of two -- change when the DM is neither cold nor collisionless. Given the importance of baryons, and that they may affect the inner dark matter distribution in satellites, we also consider different values for the subgrid parameters to explore variations in the population of satellites associated to this. To this end, we simulate cosmic structure formation in CDM, WDM and a range of SIDM cross-sections in the same $(12~\mathrm{Mpc})^{3}$ periodic volume. This allows us to focus on the same haloes in this suite of thirteen different simulations and study how the properties of their satellite systems change.  

This paper is structured as follows. Section 2 introduces the different models we use to simulate structure formation, from N-body to full-hydrodynamical realisations. Section 3 presents the methods used to measure and compare the properties of interest and the sample selection. This is followed by an overview of the changes in the properties of field haloes driven by different models. Subsequently, we shift our analysis to our sample of mass-selected haloes to investigate how their satellite populations are affected under different models. Finally, we investigate the cause behind the differences that these changes have had on the their satellite stripping and survivability.

\section{Simulations}

In this section we give an overview of the EAGLE subgrid physics used in this work and describe how we model the changes in the dark matter and baryon models.

\subsection{The code}

The EAGLE project \citep{Schaye.2015,Crain.2015} is a suite of
hydrodynamical cosmological simulations that follow the formation and
evolution of cosmic structure from $\Lambda$CDM initial conditions
assuming the cosmological parameter values from
\citet{Planck_Collaboration.2014}. They were performed using a
modified version of the P-Gadget3 code \citep{Volker.2005} that
incorporates subgrid prescriptions for the physics relevant to galaxy
formation and evolution: radiative cooling and photoheating \citep{Wiersma.2009}, star formation and evolution
\citep{Schaye.2004,Schaye.2008}, stellar feedback
\citep{Dalla_Vecchia.2012}, black hole seeding
\citep{Springel.2005,Booth.2009}, its subsequent growth and
stochastic, thermal AGN feedback. 

The values of the parameters used in
modelling these processes were set by requiring a good match to the
observed $z = 0.1$ galaxy stellar mass function, the distribution of
galaxy sizes and the amplitude of the central black hole mass {\em vs}
stellar mass relation. Once calibrated in this way, EAGLE reproduces a
number of population statistics \citep{Schaller.2015,Ludlow.2017}.

We use the calibration made for the higher mass resolution version of EAGLE to simulate structure formation in a periodic volume of $(12~\mathrm{Mpc})^{3}$. We populate it with $2 \times 512^{3}$ particles, half of which are dark matter and the rest gas particles. This corresponds to a particle mass resolution of $\sim 4\times 10^{5}$ and $\sim 8\times 10^{4} \, \Msun$, respectively. The initial conditions were generated using MUSIC \citep{Hahn.2011}.

\subsection{Baryonic physics}

An important parameter determining whether gas blowouts can flatten the density profiles of dark matter haloes in hydrodynamical simulations is the star formation density threshold \citep{Benitez-Llambay.2019}. This parameter sets the minimum density required for a gas particle to be eligible to become a star particle. The EAGLE subgrid physics uses a metallicity ($Z$) dependent term given by \citep{Schaye.2004}:
\begin{equation}
    \rho_{\rm th} = n_{\rm th, 0}\Big(\dfrac{Z}{0.04}\Big)^{\alpha} \, ,
    \label{density_threshold_equation}
\end{equation}
where $n_{\rm th, 0} = 10^{-1} \, \mathrm{cm}^{-3}$ and $\alpha = 0.64$. These values result in thresholds that are comparatively lower than other hydrodynamical simulations, e.g. $10^{2}\,\mathrm{cm}^{-3}$ in GASOLINE \citep{Zolotov.2012} or $10^{3}\,\mathrm{cm}^{-3}$ in FIRE-2 \citep{Fitts.2017}. Consequently, gas cannot accumulate in sufficient quantities at the centres of haloes to become gravitationally relevant before being blown out via supernovae feedback resulting from star formation. As a result, the EAGLE model cannot form cores through baryonic blowouts \citep{Navarro.1996a}.

Nonetheless, $\rho_{\rm th}$ is a free parameter of the subgrid physics. Indeed, star forming gas clouds in the real universe reach gas densities in excess of $10^{4} \, \mathrm{cm}^{-3}$ \citep{Lada.2009}. It is thus possible that internal structural changes that occur in the real Universe are not captured by the low values of star formation threshold used in the fiducial subgrid parameters of EAGLE. Thus, we explore how baryon-induced cores affect the satellite population of the objects with masses similar to our Milky Way by running models with higher density thresholds, setting $\rho_{\rm th}$ to a constant value of $10 \, \mathrm{cm}^{-3}$. Although this is still comparatively low than other simulations, it is large enough for gas blowouts to turn cusps into cores at the dwarf galaxy scale in EAGLE \citep{Benitez-Llambay.2019}. We refrain from using larger density thresholds as this would drastically reduce the efficiency of the thermal supernova feedback implemented in our simulations. This would make dwarf galaxies unrealistically baryon-dominated in their centres at all times \citep{Benitez-Llambay.2019}, unless other subgrid model parameters are re-calibrated. We have checked that basic galaxy properties, such as the stellar-to-halo-mass relation, do not change significantly across the models used in this work.

To distinguish between both baryonic physics models, we henceforth refer to the fiducial, low density threshold value as LT and the higher value as HT from here on. Simulations without baryons are referred to as dark matter only (DMO).

\subsection{Warm dark matter}

We obtain the power spectrum of WDM, $P_{\rm WDM}(k)= T^{2}(k)P_{\rm CDM}(k)$, using the transfer function of \citet{Bode.2001}:
\begin{equation}
    T^{2}(k) = [1 + (\alpha k)^{2\nu}]^{-5/\nu} \, .
\end{equation}
Here, $\nu$ is a fitting constant equal to 1.2 and the parameter $\alpha$ depends on the assumed mass of the WDM particle:
\begin{equation}
    \alpha = 0.049\Big[\dfrac{m_{\rm th}}{\mathrm{keV}} \Big]^{-1.11}\Big[ \dfrac{\Omega_{\rm WDM}}{0.25}\Big]^{0.11} \Big[ \dfrac{h}{0.7} \Big]^{1.22} \, h^{-1} \, \mathrm{Mpc}\, .
\end{equation}

For this work we assume $m_{\rm th} = 2.5\, \mathrm{keV}$. This is lighter than the equivalent thermal relic mass of a 7~KeV sterile neutrino model associated with the unidentified 3.5~KeV X-ray line \citep{Boyarsky.2014}. Nonetheless, we choose this value to enhance the differences with respect to CDM to allow for an easier comparison. We can estimate the mass scale where the differences with respect to CDM are noticeable, $m_{1/2}$. It corresponds to the Jean's mass of a perturbation with a wavelength equal to the one where the WDM power spectrum is half of the CDM one. For the values used in this work, $m_{1/2} = 1.4 \times 10^{9} \Msun$.

\subsection{Self-interacting dark matter}

Self-interactions are modelled using the Monte-Carlo implementation described in \citet{Robertson.2017}. Dark matter particles can scatter each other when they are closer than the Plummer-equivalent softening length of the simulations. The probability of any two neighbouring particles scattering is a function of their relative velocity and the assumed cross-section. 

In this study, we use three different cross-sections; two velocity-independent, isotropic cross-sections of 1 and 10~$\mathrm{cm}^{2} \mathrm{g}^{-1}$ and an anisotropic, velocity-dependent one given by:
\begin{equation}
    \dv{\sigma }{\Omega} = \dfrac{\sigma_{T,0}}{4 \pi \Big(1 + \dfrac{v^{2}}{w^{2}}\mathrm{sin}^{2}\dfrac{\theta}{2}\Big)^{2}}\; ,
\end{equation}
where $v$ is the relative velocity magnitude between particles in their centre of mass frame and $\theta$ the scattering angle relative to their incoming direction. The above expression results from assuming that the particles scatter in a Yukawa potential under the Borne approximation \citep{Ibe.2010}. 

The parameters $w$ and $\sigma_{T,0}$ correspond to the velocity scale below which the cross-section is roughly constant and its asymptotic, low-velocity value, respectively. We use $w = 560\,\mathrm{km}\mathrm{s}^{-1}$ and $\sigma_{T,0} = 3.04\,\mathrm{cm}^{2}\,\mathrm{g}^{-1}$ to reproduce the best-fitting mass-dependent cross-section of \citet{Kaplinghat.2016}, which is derived from constraints on the inferred cross-section from dwarf to cluster scale haloes. In practice, these values yield an approximately constant cross-section of $\sim 3 ~\mathrm{cm}^{2} \mathrm{g}^{-1}$ on dwarf galaxy scales. 

\section{Methods}
Here we discuss how we find cosmic structure and link subhaloes across snapshots to build their merger trees. We also show how we remove WDM spurious groups, select our sample of haloes and their satellites and correct for orphan galaxies. The former are satellite galaxies whose host dark matter halo has been lost from the halo catalogues. Their omission leads to underestimates of the satellite radial distributions in the central regions of haloes, where they are the dominant population.

\subsection{Structure finding and merger trees}

To identify cosmic structures, we assign particles into distinct
groups according to the friends-of-friends (FoF) percolation algorithm
\citep{Davis.1985}. Each group is made up of particles that are within
0.2 times the mean interparticle separation from one
another. Gravitationally bound substructures are found with the SUBFIND algorithm \citep{Springel.2001}, which, using particle velocity and position information, identifies self-bound structures within a larger FoF group.

We follow the time evolution of all SUBFIND groups using their merger trees, which are built by cross-matching a subset of the most bound particles between consecutive time outputs \citep{Jiang.2014}. This implementation is able to link SUBFIND groups that have temporarily disappeared from the catalogues (e.g. due to insufficient density contrast near centres of more massive haloes) for five consecutive data outputs or less. The main progenitor branch is then found by identifying the progenitor branch with the largest integrated mass \citep{DeLucia.2007}. This reduces the influence that halo switching, prone to occur during major mergers, has on the identification of the main progenitor at high redshifts.

\subsection{WDM spurious group removal}

Particle-based simulations starting from a density perturbation power spectrum with a resolved cut-off produce spurious structure along filaments. This is a consequence of the discrete representation of the underlying density field \citep{Wang.2007}. Consequently, this results in an artificially high number of objects below the mass scale where no structure is expected to form.

In this study, we remove them from the WDM simulations using the two criteria of \citet{Lovell.2014}. Firstly, we remove all groups whose peak bound mass is below the mass scale at which the number of spurious groups is equal to genuine ones, $ M_{\rm lim}$. This is related to the mass resolution of the simulation and the assumed power spectrum via:
\begin{equation}
    M_{\rm lim} = 5.05 \, \bar{\rho} d k^{-2}_{\rm peak}, 
\end{equation}
where $d$ is the mean interparticle separation, $k_{\rm peak}$ the wavelength at which the dimensionless power spectrum, $\Delta^{2}(k)$, peaks and $\bar{\rho}$ the mean density of the universe. For the simulations and WDM model used in this study, $M_{\rm lim} = 1.4 \times 10^{8} \, \Msun$ .

Finally, we select the particles bound to the group when it first reached half of its peak bound mass. We then compute the inertia tensor of those particles in the initial conditions and define the sphericity as the ratio between the smallest and largest eigenvalue of their inertia tensor, $s \equiv c/a$. All groups with $s \leq 0.16$ are removed, since the Lagrangian regions associated to spurious groups are significantly more flattened than those in which genuine haloes form. 

\subsection{Halo and subhalo matching across simulations}

We match the main SUBFIND group of each FoF group across simulations by selecting their 100 most bound particles as identified by SUBFIND. We then select a candidate match by identifying which group the majority of the particles belong to in the other simulations. The process is then repeated in reverse, and if this bijective process is successful, we confirm the match. 

Matching substructure is less trivial owing to the fact that the same object may have followed different paths and have been stripped to varying degrees once it entered the virial region of a larger object. To minimise the effect of these differences, we perform the bijective match at the time when their bound mass peaked.  

\subsection{Sample Selection}

As we are interested in studying the satellite system of haloes similar to that of our own Milky Way, we restrict our analysis to haloes of mass
$M_{200}$\footnote{$M_{200}$ is defined as the mass contained within a sphere of mean density 200 times the critical density of the universe.} at $z = 0$ in the range $0.5 -2.5\times 10^{12} \,
\Msun$. This is within a factor of two from recent observational estimates of the Milky
Way's halo mass \citep{Callingham.2019,Cautun.2020}. Eight
haloes satisfying this criterion were identified in each version of the simulations. However, one is undergoing a merger at $z=0$, which we remove from further consideration.

Their resolved satellite systems are defined by identifying all SUBFIND groups that are within 300~kpc from the centre of their host halo and have one or more bound stellar particle at $z = 0$. We also enforce that the identified structures are heavily dark matter dominated, namely, $M^{\rm DM}_{\rm SUB}/M^{\rm tot}_{\rm SUB} > 0.8$. This additional condition stems from the fact that dense clumps of gas in the HT versions are identified as self-bound structures by SUBFIND. Their inclusion in the satellite population would lead to biased radial distribution functions, as they form in the inner few kiloparsecs of the dark matter halo, where the gaseous disk is located. Some gas clumps are also present in the low threshold versions, but are far less common than in the higher density threshold counterparts.

\subsection{Orphan galaxies}

In simulations of structure formation with limited resolution, substructure can be artificially disrupted. Substructure is lost whenever its mass drops below the 20 particle threshold limit imposed by SUBFIND on bound structures. The decrease in the bound number of particles can occur, for example, when a subhalo has been tidally stripped or the density contrast is insufficiently high for it to be detected near the central regions of a more massive neighbour. This does not necessarily imply that they have been disrupted, since increasing the particle mass resolution would lead to their ongoing survival for a longer time. This is both due to an increased capability in tracking objects to lower masses, as $m_{\rm limit} \sim 20 m_{\rm dm}$, and due to a reduction in the effect of tides resulting from smaller artificial cores.

Thus, accounting for these `disrupted' objects improves the convergence of the predicted radial distribution function of satellites around Milky Ways \citep{Newton.2018}. Moreover, they are required to correctly predict the satellite luminosity functions at stellar masses below $10^{5} \, \Msun$, even in high resolution simulations \citep{Bose.2020}. 

In this study, we tag as orphans all dark matter haloes that had at least one bound stellar particle before being lost from the merger trees. We then use their most bound DM particle -- identified during the last data output when they were resolved -- as a proxy for the position and velocity of the orphan galaxy. A small subset of orphans end up sharing the same tracer particle ID. In such cases, we discard the higher redshift counterparts and keep the one orphaned at a later time.

Once the orphan population is identified, we track their positions until one of the two conditions given in \citet{Simha.2017} are fulfilled. The first one is that they have existed for longer than the time for their orbit to decay due to dynamical friction: 
\begin{equation}
    \dfrac{T_{\rm df}}{\tau_{\rm dyn}} = \Big(\dfrac{r}{r_{\rm circ}}\Big)^{-1.8}\Big(\dfrac{J}{J_{\rm circ}}\Big)^{0.85}\dfrac{M_{\rm FoF}(<r)/M_{\rm sub}}{2B(1)\ln \Lambda} \, ,
\end{equation}
where $r$ and $J$ are the orbital radius and angular momentum of the orphan, and the corresponding values for a circular orbit of the same binding energy are $r_{\rm circ}$ and $J_{\rm circ}$, respectively. The Coulomb logarithm is taken to be $\ln \Lambda = \ln M_{\rm vir}/M_{\rm sub}$ and $B(x) \equiv \mathrm{erf}(x) - 2x e^{-x^2}/\sqrt{\pi} $. The dynamical timescale of the halo, $\tau_{\rm dyn}$, is estimated as:
\begin{equation}
    \tau_{\rm dyn}(z) = \dfrac{1}{\sqrt{4\pi G \Delta_{\rm vir}(z)\rho_{\rm crit}(z)}}\, ,
\end{equation}
where $\rho_{\rm crit}$ is the critical density of the universe and $\Delta_{\rm}$ is the overdensity of a just-collapsed spherical top hat density perturbation \citep{Bryan.1998, Eke.1996}: 
\begin{equation}
    \Delta_{\rm vir}(z) = 18\pi^{2} + 82[\Omega_{\rm m}(z)-1] + 39[\Omega_{\rm m}(z)-1]^2 \, .
\end{equation}
The dynamical friction timescale is first calculated immediately after the galaxies are orphaned. If the orphan subsequently enters the virial region of a more massive FoF group, we re-calculate and update its value.

The second condition is to stop tracking orphans once they come within a radius that encloses a mean density equal to the mean density of the orphan, $\bar{\rho}_{\rm FoF}(<R_{\rm tid}) = \bar{\rho}_{\rm sub} (<R_{\rm sub})$. For the spatial scale of the subhalo, $R_{\rm sub}$, one may chose $R_{\rm max}$ or the half-light radius of the galaxy it hosts, $R_{50}$. Here we use the latter, since we are interested in modelling when the luminous component of the galaxy is affected by tides. A subset of orphans have no associated $R_{50}$, e.g those with only one bound stellar particle. In such cases, we compute the median of $\rho(<R_{50})/\rho(<R_{\rm max})$ for orphans with known $R_{50}$ and multiply $\rho(<R_{\rm max})$ by this correction factor to estimate $\rho(<R_{\rm 50})$. 

\subsection{Orbit integration}

The typical time resolution between consecutive data outputs for our simulations ($\sim 300~ \mathrm{Myr}$) is much larger that the dynamical timescales of the central regions of the haloes in the mass range we study here. This means that outputs are unlikely to `catch' satellites near pericentre, potentially leading to an underestimate in their numbers in the central regions. This can affect estimates for the central radial distribution of satellites, as well as whether the tidal disruption criterion is fulfilled. 

We interpolate the orbits of satellites between consecutive data outputs. Here we use the method described in \citet{Richings.2020}, with a few notable differences. Firstly, we use the AGAMA package \citep{Vasiliev.2019} instead of GALPY \citep{Bovy.2015}. Secondly, we align the z-axis of the coordinate system with the $z=0$ angular momentum of the galaxy's stellar component, if present. Finally, we use an axisymmetric multipole expansion for the potential sourced by the DM and a cylindrical one \citep{Cohl.1999} for that of the baryons. The latter choice is made to model more accurately a flattened potential. 

\begin{figure}
    \centering
    \includegraphics{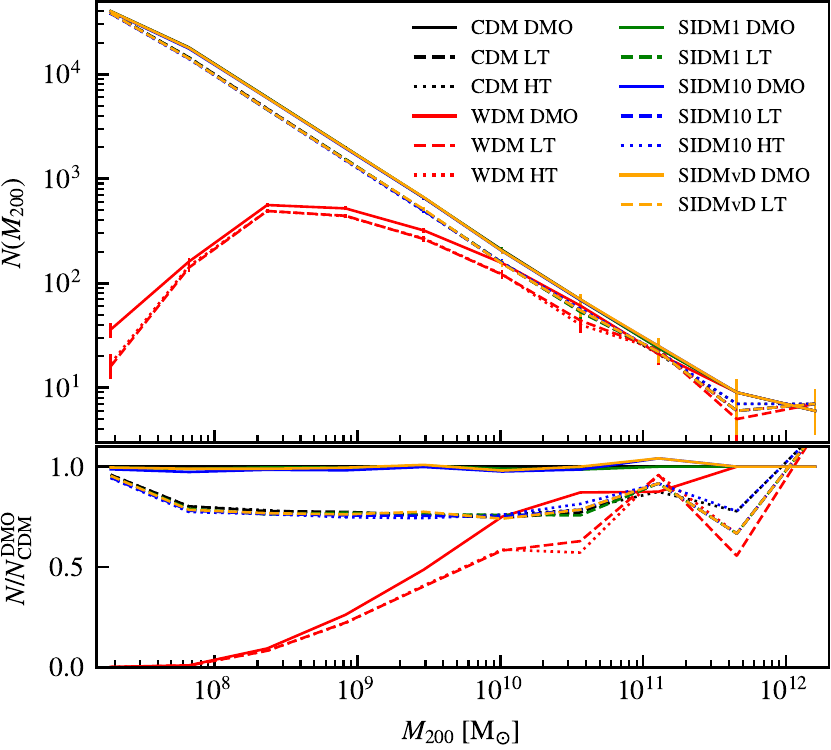}
    \caption{\textit{Top panel}: halo mass functions for the CDM (black), WDM (red) and SIDM (green, blue and orange) versions of the simulation box used in this work. The line styles show whether they are measured in the DMO (solid), reference hydrodynamical (dashed) or high threshold hydrodynamical versions (dotted). The error bars correspond to the Poisson noise in each mass bin, which is largest at high masses. \textit{Bottom panel}: ratio of the halo mass function of each version relative to that measured in the CDM DMO version.}
    \label{halo_mass_function}
\end{figure}

\section{Field haloes}
Here we discuss how the global and internal properties of field haloes differ among different DM models, as well the choice of subgrid physics. We begin by comparing the abundance and global properties of all haloes, luminous or dark, across our simulations. We discuss changes in the galaxy formation efficiency, namely, the fraction of luminous haloes in a given mass range. Finally, we study how the dark matter distribution differs between matched pairs of DM haloes across all simulations.

\subsection{Halo mass functions}

In Fig.~\ref{halo_mass_function} we show the halo mass function as measured in all simulations available for this volume. We have defined the virial mass as the mass contained within a sphere whose mean density is 200 times $\rho_{\rm crit}$. Focusing on the CDM DMO version, we show the expected power law dependence on $M_{200}$ in the mass range $10^{8} - 10^{11} \Msun$. At higher masses, we observe a deviation from this behaviour. This is driven by Poisson fluctuations that arise as a consequence of the small number of massive objects in our simulations. Indeed, within a volume of $(12\,\mathrm{Mpc})^3$, we expect less than 10 MW-mass haloes to form.

\begin{figure}
    \centering
    \includegraphics{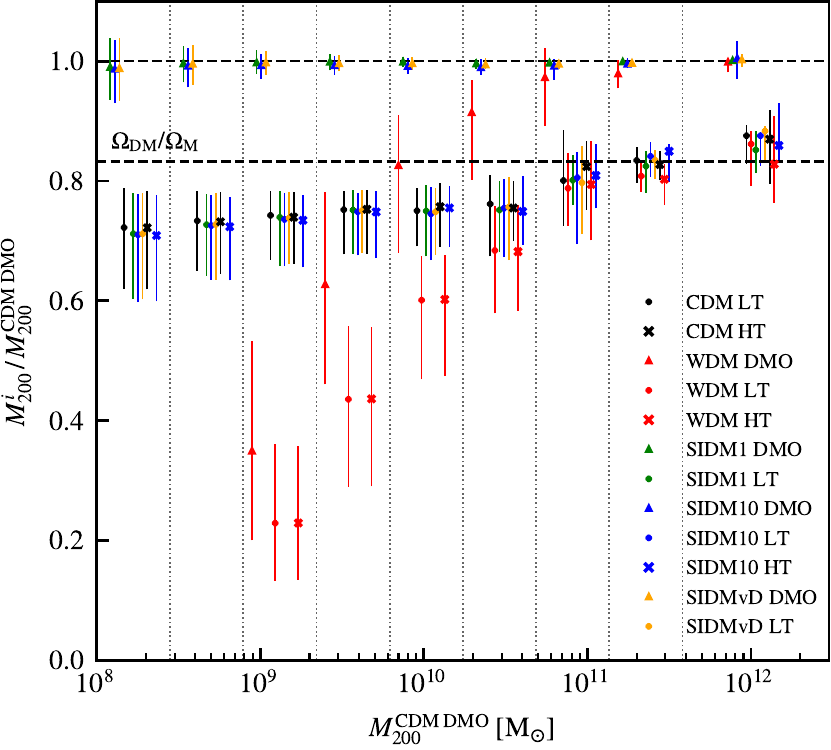}
    \caption{Median virial mass ratio of all field haloes relative to their CDM DMO counterparts, measured at $z=0$ and binned as a function of the $M_{200}$ of the CDM DMO counterpart. This is shown for the CDM (black), WDM (red) and SIDM (green, blue and orange) models, as indicated by the legend. The symbols indicate whether the simulation is DMO (triangle), LT (circle) or HT (cross). The error bars show the 16th and 84th percentiles, with the symbols being offset with respect to each other for clarity. The vertical dotted lines indicate the mass ranges used to bin haloes and the horizontal lines the unity ratio and the universal dark matter fraction.}
    \label{mass_ratio}
\end{figure}

The corresponding SIDM DMO simulations show no appreciable differences relative to the CDM versions in the sampled mass range, regardless of the cross-section value. This is because the primordial density fluctuation power spectrum was assumed to be the same across these two models. The addition of self-interactions primarily affects the central regions of DM haloes, where higher densities allow for more frequent interactions between particles. There are no significant differences in the distribution of dark matter near the virial radius nor in the number of objects that form, and hence there are no changes in the halo mass functions relative to CDM. 

On the other hand, the WDM DMO simulation shows large differences with respect to the CDM and SIDM models. Although at higher masses these are negligible, they become significant close to and below the half-mass mode of our WDM model. This is evident as a reduction in the number of haloes at a fixed $M_{200}$ on those mass scales. This is due to the suppression of small spatial scale density perturbations, which results in fewer low-mass objects forming compared to the CDM and SIDM models. However, we point out that the systems that do form are less massive than their CDM counterparts, as shown in Fig.~\ref{mass_ratio}.

In this mass range, the hydrodynamical versions of all models exhibit a systematic suppression with respect to their DMO counterparts. This is a consequence of the loss of baryons within the virial region of haloes at early times, which induces a shift in the halo mass functions towards lower masses. As shown in Fig.~\ref{mass_ratio}, all models in the largest mass bins have ratios close to the universal dark matter mass fraction. The mass loss is entirely explained by the removal of a large fraction of the baryons by feedback at early times. Focusing on lower masses, we see that the ratio for the CDM and SIDM models approaches a constant fraction that is lower than $\Omega_{\rm DM}/\Omega_{\rm m}$. This because the loss of baryons at early times hinders subsequent mass growth due to the resulting shallower gravitational potential well, leading to overall less massive haloes \citep{Sawala.2013}. The case of WDM is the combination of the above together with a mass decrease arising from the cut-off in the power spectrum.

\begin{figure}
    \centering
    \includegraphics{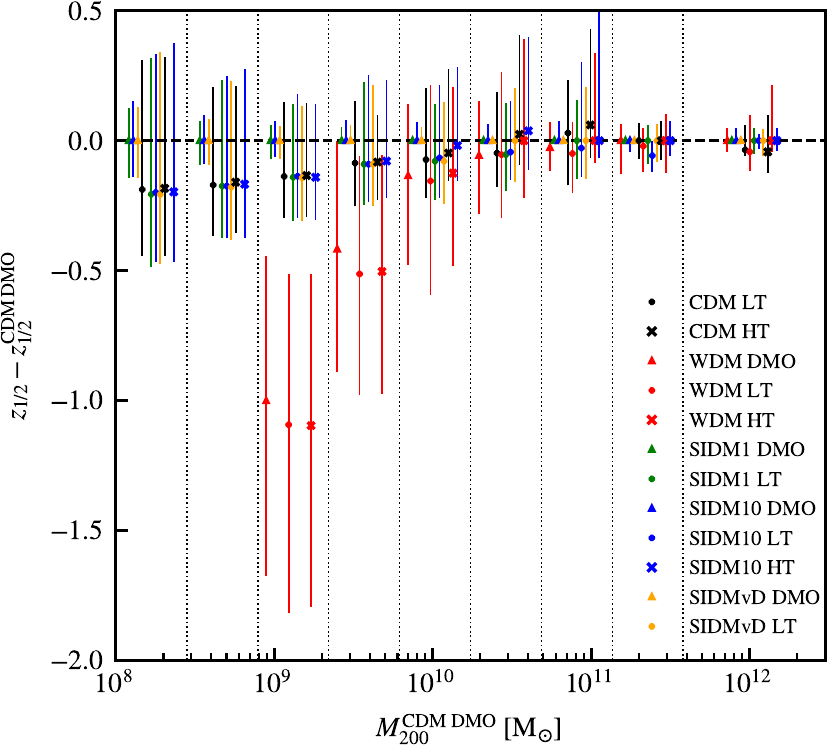}
    \caption{Difference in the formation time of haloes relative to their CDM DMO counterparts, quantified by the redshift at which the halo reached half of its $z = 0$ virial mass. This is shown for the CDM (black), WDM (red) and SIDM (green, blue and orange) models, as indicated by the legend. The symbols indicate whether the simulation is DMO (circle), LT (cross) or HT (triangle). The error bars show the 16th and 84th percentiles of the distribution in each mass bin, with the symbols being offset with respect to each other for clarity. }
    \label{formation_time}
\end{figure}

\subsection{Halo formation times}

The epoch at which haloes form determines the shape and normalisation of their DM density profiles. This is because the formation time reflects the density of the Universe when the density perturbation decoupled from the Hubble flow. As discussed above, the early loss of baryons can slow down the mass growth of DM haloes. Moreover, a cut-off in the power spectrum can also delay the formation of haloes. 

To explore how the differences made to our models alter the formation time of haloes, we compare how formation times vary across matched pairs, relative to their CDM DMO counterparts. For this purpose, we identify the formation time with the redshift at which the main progenitor first reached half of its $z=0$ virial mass, $z_{1/2}$. We compare how the median and scatter of this ratio varies as a function of mass in Fig~\ref{formation_time}.

We first note the similarities between the DMO versions of SIDM and CDM across the mass range studied here. Again, this is because the power spectra of inital density perturbations were the same in both models. On the other hand, the WDM DMO counterparts exhibit a formation delay that increases towards lower masses. This means that they form when the Universe is less dense compared to haloes that collapse earlier. Thus, their concentrations are expected to be lower in WDM than in their CDM and SIDM counterparts.

The hydrodynamical versions of all simulations have equal formation times for larger mass haloes, but begin to form slightly later at lower masses. This is caused by the loss of baryons at early times, which results in a measurable slow down in the growth rate of the halo due to the shallower potential well. This leads to lower virial masses at $z=0$ relative to their DMO counterparts, as discussed previously.

The above changes in the formation times of haloes have important implications on the fraction that host galaxies. This is because the interplay between their mass accretion histories and the mass required to trigger the gravitational collapse of gas largely determines whether a halo is luminous or not at $z=0$. Thus, delayed formation times and slower growth -- indirectly probed by our $z_{1/2}$ metric -- can reduce the amount of luminous haloes in a given mass range.

\begin{figure}
    \centering
    \includegraphics{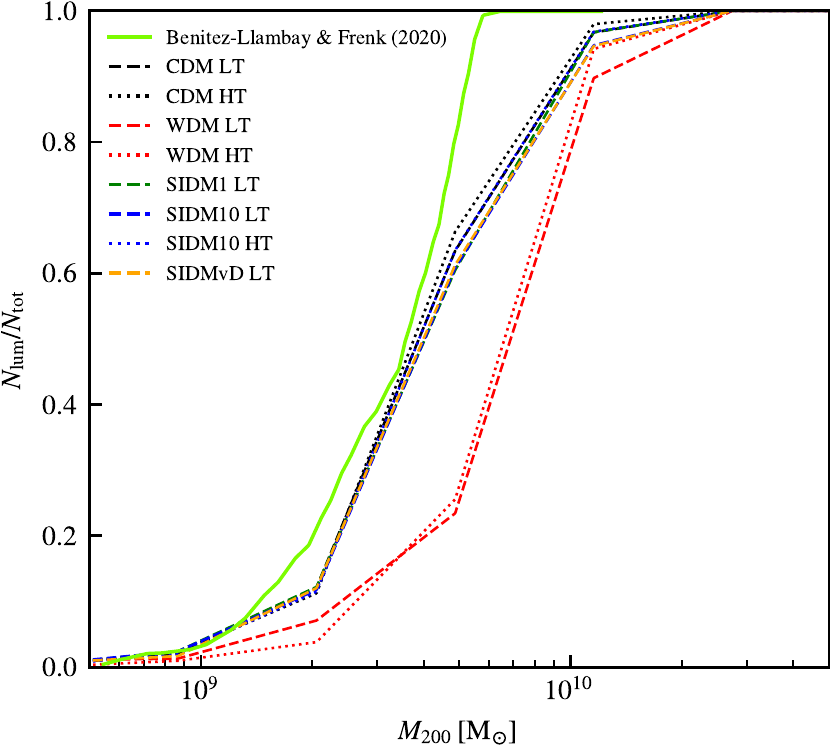}
    \caption{Fraction of field haloes that host a luminous component at $z=0$, as a function of their virial mass. This is shown for the CDM (black), WDM (red) and SIDM (green, blue and orange) models, as indicated by legend. The linestyles indicate whether the simulation assumes a low (dashed) or high threshold (dotted) for star formation. The predicted halo occupation fraction of \citet{Benitez-Llambay.2020} is shown as a green line.}
    \label{occupation_fraction}
\end{figure}

We explore this in Fig~\ref{occupation_fraction}, which shows how the halo occupation fraction (HOF) varies across different models. First, focusing on the CDM LT version, we observe three distinct regimes. At masses below $\sim 10^{9}\Msun$, no haloes host luminous components, whereas all haloes are luminous above $\sim 10^{10}\Msun$. The mass range between both limits is populated both by luminous and starless haloes.

The shape of the HOF is well understood from simple assumptions about when galaxy formation is triggered \citep{Benitez-Llambay.2020}. Essentially, any halo more massive than a redshift-dependent mass threshold, defined by the scale at which gas is unstable to gravitational collapse, will host a galaxy by $z=0$.. Before reionisation, this threshold is determined by atomic hydrogen cooling; after reionisation it is determined by the thermal state of the intergalactic gas. At high masses, all have crossed this threshold, hence all are luminous. Those at intermediate masses will cross it (or not) depending on their mass assembly histories, which vary across haloes. At lower masses, objects have not been able to trigger the gravitational collapse of gas and thus remain starless. 

The predicted (CDM) HOF of \citet{Benitez-Llambay.2020} is shown in Fig~\ref{occupation_fraction}; its midpoint agrees well with our simulations. Nonetheless, there are some differences on the high and low mass ends. These are largely driven by the binning scheme we require to measure the HOF in our simulations, which is not fine enough to capture the sharp transition. However, all haloes above $6\times10^{9} \, \Msun$ should host a galaxy, but we find some that remain starless in our simulations. We attribute this to resolution effects: the limited resolution of our simulations (a factor of 8 coarser than that in the simulations of \citealt{Benitez-Llambay.2020}) is not enough to follow accurately the rate at which the gas becomes denser as it approaches the threshold for star formation.

\begin{figure*}
    \centering
    \includegraphics{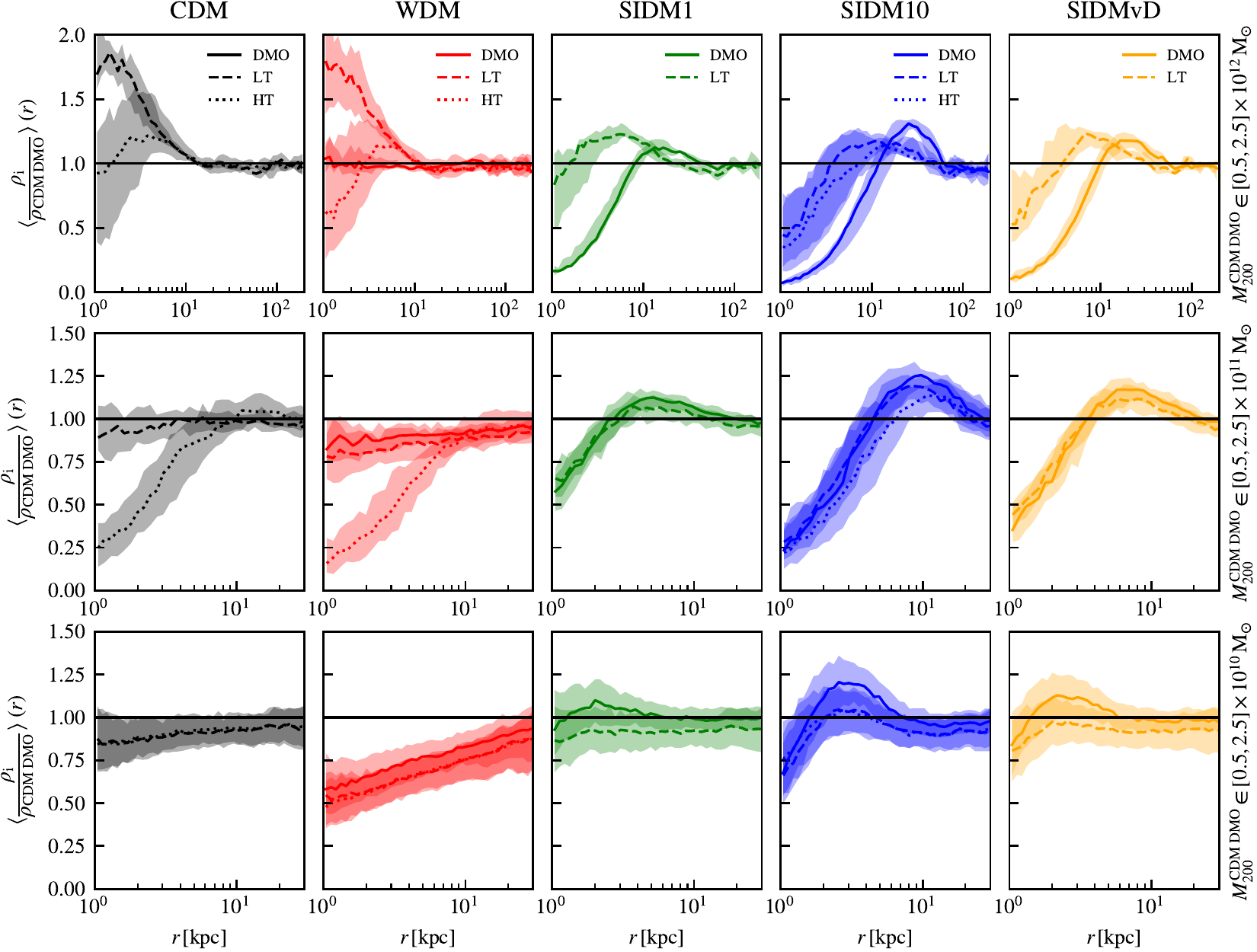}
    \caption{Median dark matter density profiles of field haloes as function of physical radius, relative to their matched CDM DMO counterparts. We use physical distances in kpc instead of $R_{200}$ to prevent the location of overdensities due to substructure changing their positions between hydrodynamical and DMO counterparts. This would occur because the virial radii of haloes change across simulations, e.g Fig.~\ref{mass_ratio}, despite substructure being located at the same physical distance from the centre. The shaded areas around each line correspond to the 16th and 84th percentiles of the distributions.  These were calculated for different mass bins in $M^{\rm CDM \; DMO}_{200}$, with each row corresponding to the range indicated on the right hand side. Note the change in the x-axis radial scale between the top and bottom rows. Different dark matter models are shown in each column, with line styles indicating whether they were measured in a DMO (solid), LT (dashed) or HT (dotted) simulations.}
    \label{density_ratios}
\end{figure*}

Turning back to the HOF measured in our simulations, we can understand the differences between all models. For the SIDM and CDM cases, regardless of the hydrodynamical model, no significant differences exist in the assembly of matched counterparts. Thus, haloes that form galaxies in one simulation always do so in the alternative models. On the other hand, the WDM simulations show a clear difference with respect to the latter two. This is connected to their delayed formation.

To understand why this is the case, consider a CDM halo that only just crosses the mass threshold for galaxy formation. Its WDM counterpart will, at a fixed redshift, 
be less massive due to its delayed formation. Consequently, it will not be massive enough to trigger the gravitational collapse of gas and will remain starless.

Evidently, the details of this simplified explanation change once a more realistic picture is considered. For example, lower concentrations of low mass haloes alter the hydrostatic equilibrium profile of gas, although this effect is minor. Moreover, the properties of reionisation also change as a reflection of the suppression of low-mass structure \citep{Yue.2012, Dayal.2017}. Finally, we note that the properties of the subset of starless haloes which retain their gas content after reionisation (reionisation limited HI clouds, \citealt{Benitez.Llambay.2017}), remain as of yet, unexplored. It would be interesting to contrast how their properties and abundance compare to those formed in CDM, potentially yielding additional constraints on the nature of DM.

\subsection{Density profiles}

An important prediction of simulations of cosmic structure formation is the spherically-averaged radial density profile of DM haloes. Their profiles in CDM DMO simulations are quasi-universal over 20 orders of magnitude in halo mass \citep{Wang.2020} and well described by the NFW \citep{Navarro.1996b} and Einasto \citep{Einasto.1965} formulas, which predict centrally divergent cusps. However, we expect significant changes to the the internal structure of DM haloes in the different models we study. Examining how they differ is an important step in understanding differences in the predicted $z=0$ satellite system, since it influences how strongly they are tidally stripped \citep{Penarubia.2010}.

Firstly, low-mass WDM haloes are likely to be less concentrated resulting from the delay in their formation relative to CDM. Scattering due to self-interactions will drive the centre of an initially cuspy profile to an isothermal, constant density core \citep{Rocha.2013,Robertson.2021}. Finally, the inclusion of baryons and different subgrid prescriptions may cause additional differences such as cores in CDM and WDM haloes, contraction of high mass haloes and an overall reduction in the DM density due to delayed growth.

We study in Fig.~\ref{density_ratios} how different choices for the DM model and baryonic physics alter the density profiles in three different mass bins, $\mathrm{M}_{200} \in [0.5,2.5]\times 10^{12}\Msun$, $\mathrm{M}_{200} \in [0.5,2.5]\times 10^{11}\Msun$, and $\mathrm{M}_{200} \in [0.5,2.5]\times 10^{10}\Msun$. The latter corresponds roughly to the least massive haloes still able to form galaxies (see Fig. \ref{occupation_fraction}). 

We have matched all central haloes in these bins to their CDM DMO counterparts. We then estimate their densities using logarithmically spaced spherical shells in physical distance and express them relative to the density of their CDM DMO counterparts. Finally, we average across all haloes in these mass bins that satisfy all three relaxation criteria of \citet{Neto.2007}:
\begin{itemize}
    \item The virial ratio $|2K/U|$ should be less than 1.35.
    \item The centre of mass, measured using all DM particles within the virial region of the halo, should be within 0.07$R_{\rm vir}$ from the centre of potential.
    \item The substructure mass fraction should be less than 10\%.
\end{itemize}

Focusing first on the high mass haloes, we see large differences across different DM and baryonic physics models. For CDM, the addition of baryons has no effect on the DM density at large radii. At smaller radii, there is an $\sim 80\%$ enhancement in the DM density in the LT version. The origin of this is the contraction of the halo in response to the formation of the galaxy at its centre. The HT simulation shows differences in the central parts, most notably a lower median density ratio. Nonetheless, it is still consistent with the LT one beyond $\sim 2~\mathrm{kpc}$ within the 1$\sigma$ scatter. We have examined the profiles individually and note some of the lowest mass HT haloes within this mass bin have cores, whereas more massive ones have similar (or greater) contractions relative to their LT counterparts. These differences arise because the properties of the stellar component have changed across simulations, e.g. their masses, sizes and if they form a bar, its dipole moment strength. This affects how much the bar torques the surrounding dark matter \citep{Forouhar.2022}.

The differences in SIDM relative to CDM depend on the assumed cross-section, although they are all have lower central densities. Identifying the core radius with the radius at which the density ratio first crosses unity, we see it increases monotonically with the particle cross-section. This occurs because the radius at which the profiles become approximately isothermal depends on the scattering rate of particles, and thus on their cross-sections. The removal of DM from the centre to intermediate radii causes a localised enhancement whose magnitude and location sensitive to the assumed SIDM cross-section. Thus, parametric density profiles fitted to the central density, such as the generalised NFW \citep{Zhao.1996}, do not fit these haloes well.

The addition of baryons in SIDM reduces the differences in the central regions of haloes compared to CDM DMO. For example, the median core radius decreases from 8 and 13~kpc to 1.5 and 4~kpc, for SIDM1 and SIDM10, respectively. The enhancement in density at intermediate radii becomes more similar across SIDM models. Both are a consequence of the interplay between two competing effects: self-interactions driving a decrease in density and halo contraction counteracting it.

Focusing on the lower masses, the CDM hydrodynamical simulations produce profiles that are consistently less dense than their DMO counterparts. Although the offset relative to CDM remains constant throughout the radial ranges shown, there is a slight dependence on radius. We attribute both differences to the early loss of baryons and subsequent delay in formation time, which leads to lower densities and concentrations. The similarity between the LT and HT simulations is due to the baryonic component of these galaxies being small. Thus, baryonic blowouts are not able to perturb the inner dark matter distribution. 

Both the DMO and hydrodynamical WDM simulations show a much stronger radial offset relative to CDM. This is because they are significantly less concentrated than their CDM counterparts. Thus, their shapes are very different.

\begin{figure*}
	\centering
	\includegraphics{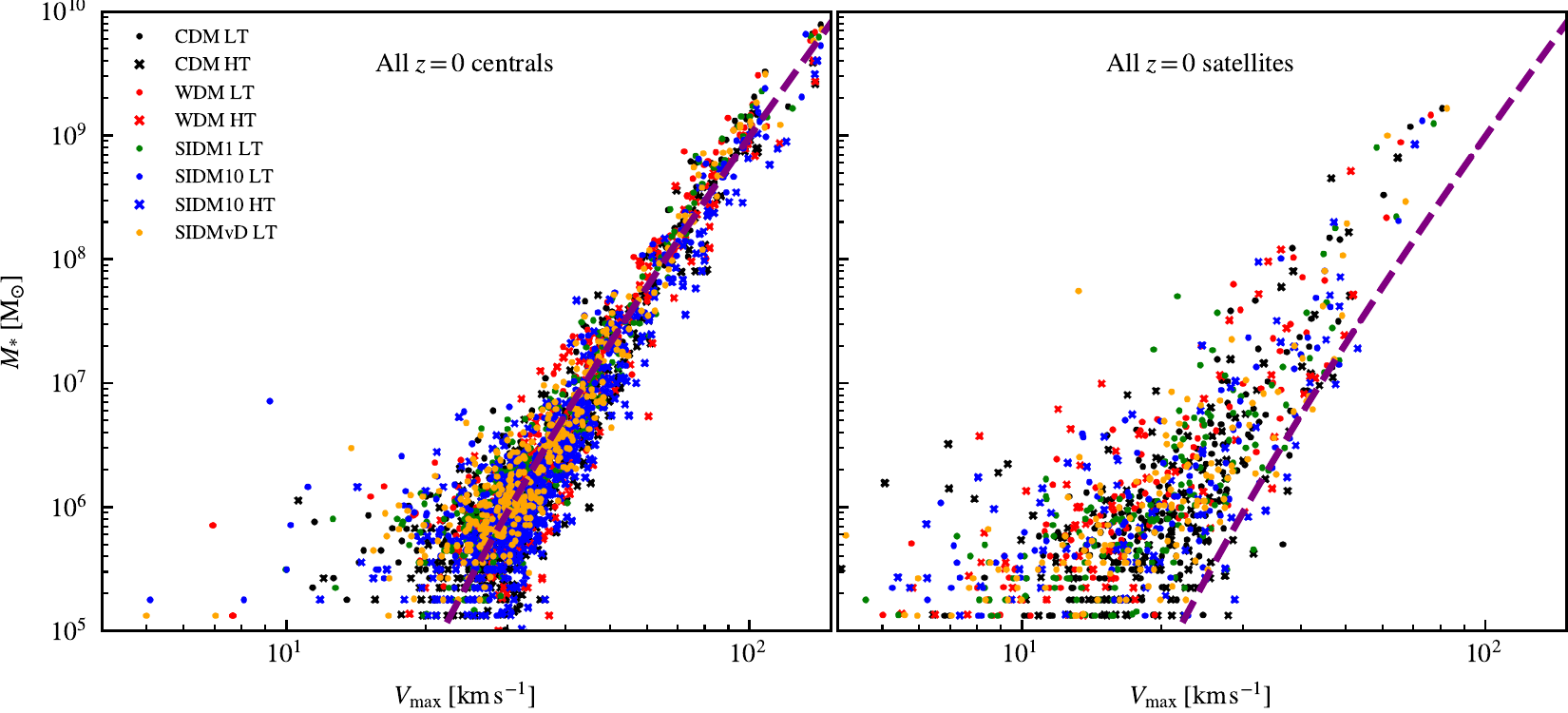}
	\caption{\textit{Left panel}: relation between the stellar mass -- within a spherical aperture of 30~$\mathrm{kpc}$ -- and $V_{\rm max}$ of all central galaxies, with each marker corresponding to different DM and baryonic physics versions, as per the legend. The purple dashed line is the best-fit relation of the form $M_{*}\propto V^{\gamma}_{\rm max}\exp{-V_{\rm max}^{\nu}}$ to the CDM LT distribution, which is similar across all simulations considered in this work. \textit{Right panel}: bound stellar mass of all $z = 0$ satellites of the studied halo sample across all simulations, as a function of their maximum circular velocity. The purple line corresponds to the same as in the left panel.}
	\label{Mstar_Vmax_relation}
\end{figure*}

Finally, low-mass SIDM DMO haloes show differences relative to CDM DMO similar to their more massive counterparts: a decrease in the central density and an enhancement at intermediate radii. Although it might seem as if the density suppression is less severe than for the high mass haloes, this is because the radial scale (relative to the virial radius), is not the same in the bottom and top panels. Once again, the inclusion of baryons reduces the differences relative to CDM. In fact, the lowest cross-section of $1~\mathrm{cm}^{2}\mathrm{g}^{-1}$ shows no significant changes within the radial range shown. At large radii, we have the same constant density observed in CDM and WDM. 

\section{Satellite systems}

As discussed in the previous section, changes to the DM model and baryonic physics lead to differences in the overall number of haloes that form, their internal structure and the fraction that host galaxies. We now focus on how these changes propagate to the satellite population of haloes.
We begin with a comparison of the $z=0$ properties, followed by a detailed analysis of the main causes for the differences. Finally, we also consider corrections to account for orphaned galaxies, which are the dominant population in the central tens of kiloparsecs and belong to the ultra-faint regime.

\subsection{A first look at the effect of tides}

The left panel of Fig.~\ref{Mstar_Vmax_relation} shows the stellar mass to $V_{\rm max}$ relation measured for all central galaxies at $z=0$. We see no systematic differences between models within the scatter, although the stellar components in HT can be slightly less massive than those in LT. Nonetheless, the best fit power law model with an exponential truncation, $M_{*} \propto V^{\gamma}_{\rm max}\exp{-V^{\nu}_{\rm max}}$, is similar in all of them. This fit was done using galaxies with $V_{\rm max} > 30 \mathrm{km} \, \mathrm{s}^{-1}$ and $M_{*} > 10^{6}\Msun $ to exclude heavily-stripped backsplash haloes, which are significantly offset from the mean relation, and galaxies with less than ten stellar particles. 

The equivalent relation for all the $z=0$ satellites is shown in the right panel of the same figure. The observed offset at fixed stellar mass with respect to the relation for the centrals reflects the effects of tidal stripping. These remove mass as the satellites orbit more massive objects, decreasing their $V_{\rm max}$ over time. This primarily affects the DM, which occupies the less bound outskirts of the halo. The stellar component remains undisturbed for much longer than the DM, since it is more centrally concentrated. 

\subsection{Stellar mass functions}
\begin{figure}
    \centering
    \includegraphics{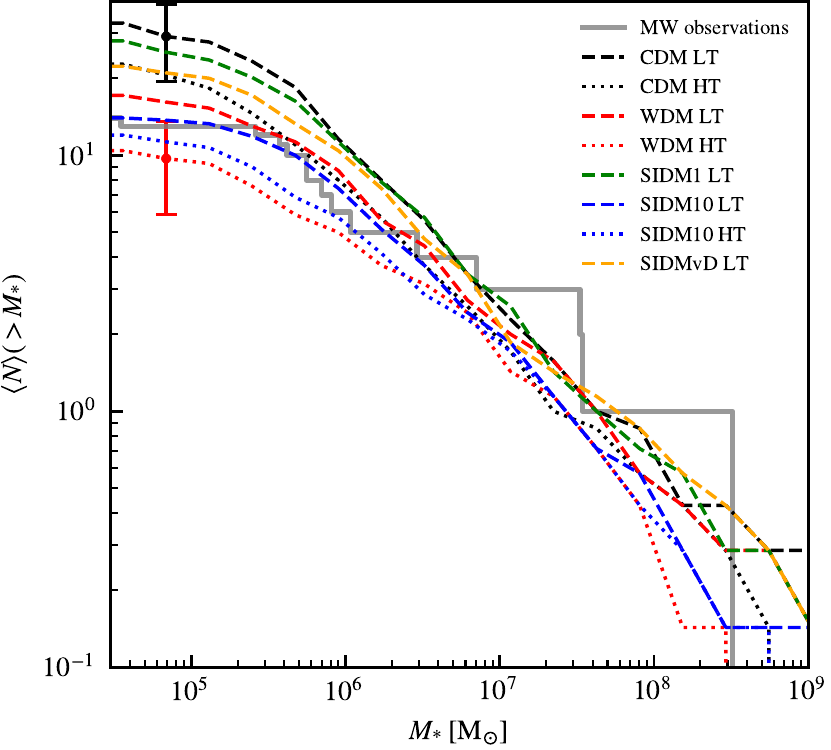}
    \caption{Stellar mass functions of the $z=0$ resolved satellites, averaged across all our sample in a given simulation. The colours encode the dark matter model used, with the line styles showing the baryonic physics employed, as indicated in the legend. The error bars indicate the standard deviation of the CDM LT and WDM HT distributions at $M_{*} \sim 10^{5} \Msun$, which is roughly 35\% of their values. This is also true for all other models in this mass range. The inferred stellar mass function based on observations is shown by the grey stepped line. These were calculated by taking the $L_{V}$ of \citet{McConnachie.2012} and applying a mass-to-light ratio correction of 1.6 to all satellites except for the LMC (for which we use $M_{*}/L_{V} = 0.7$; \citealt{Woo.2008}).}
    \label{mstar_function}
\end{figure}

In Fig~\ref{mstar_function} we show the cumulative distribution of stellar mass for our sample of haloes. These were measured by selecting all satellites, using their SUBFIND bound stellar mass and averaging across all haloes each simulation. We only show the mass regime resolved by our simulations, which corresponds to masses larger than those of the ultra-faint satellite population. Nonetheless, it is clear that the number of $z=0$ satellites above $M_{*} = 10^{5} \Msun$ is strongly dependent on the assumed DM and baryonic physics model. 

The most numerous populations occur in the CDM LT simulation, as expected. This is because their haloes are cuspy and more concentrated than in all of the other hydrodynamical models. Thus, they are more resilient to tides. When the density threshold for star formation increases (HT) -- and gas blowouts are able to carve cores -- the number of satellites decreases by about a third. This is evidence for increased stripping in the profiles with profiles.  

The SIDM simulations also show a reduction in numbers that increases monotonically with the cross-section. As we saw in the previous section, SIDM models form the same number of haloes and galaxies as CDM. The only difference is the number of satellites that survive to $z=0$. Therefore, the lack of satellites relative to CDM indicates an stronger stripping and destruction due to central cores. As we saw in the previous section, the cores driven by the DM self-interactions become larger when the cross-section is larger.

Finally, the WDM satellite population is less numerous than in CDM. Contrary to SIDM, a simple interpretation on what causes the suppression is less trivial. WDM forms fewer haloes and galaxies, but lower concentrations may also play a role in exacerbating the suppression \citep{Bose.2017}. We discuss this in more detail in the following section, where we estimate how important each of these effects are. As in CDM, the increase in the density threshold for star formation leads to enhanced suppression relative to their LT counterparts.

The shape of the stellar mass functions are similar in all models at the higher stellar mass end, but there are large differences at lower stellar masses. Nonetheless, the most similar models to CDM LT (SIDM1 and SIDMvD) only show significant differences at $\sim 10^{6}\,\Msun$ whereas the other models already exhibit them at $\sim 10^{7}$ to $10^{8}\,\Msun$

\begin{figure}
    \centering
    \includegraphics{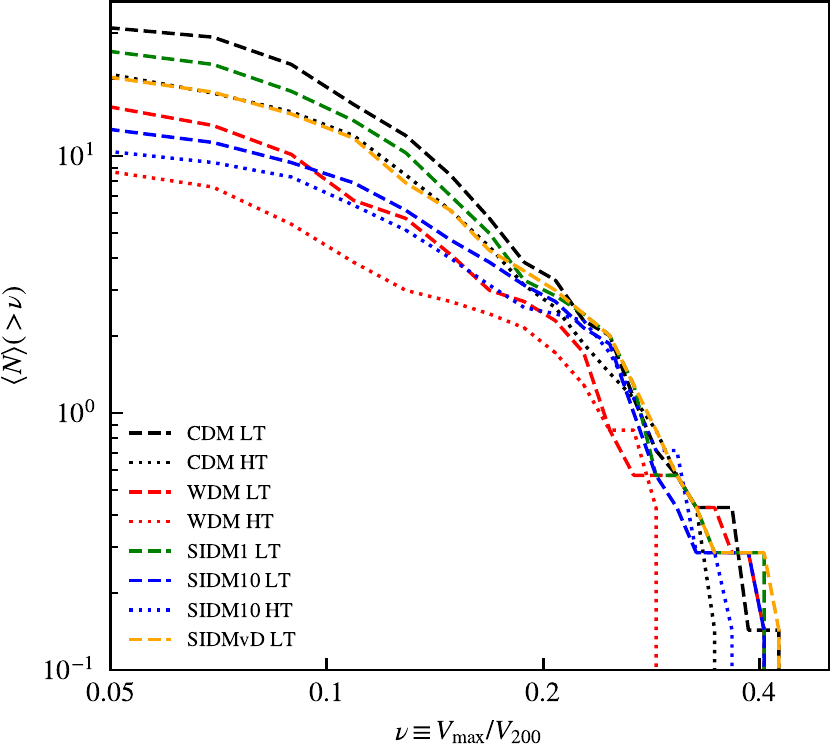}
    \caption{Distribution of the maximum circular velocities for all resolved satellites at $z=0$, relative to the $V_{200}$ of their host halo. The colours encode the dark matter model used, with the line styles the baryonic physics employed, as the legend indicates.}
    \label{averaged_vmax_distribution}
\end{figure}

We show the observed stellar mass functions for the MW satellite population as a grey stepped line in Fig~. \ref{mstar_function}. It is worth noting that the MW satellite population beyond the eleven classical satellites is incomplete. This is because surveys used for most discoveries below the classical satellite mass regime -- SDSS \citep{2015.ApJS} and DES \citep{Bechtol.2015, Drlica-Wagner.2015} -- are limited to certain regions of the sky and are flux-limited. Thus, the observational data should be considered a lower bound to the number of satellites. Nonetheless, the correction for incompleteness, based on the assumption of a CDM-like radial distribution, amounts to just a few satellites in the range shown here \citep{Newton.2018}. 

When compared to observations, the \textit{average} satellite stellar mass functions of our haloes predict more or fewer low-mass satellites above $M_{*} = 10^{5}\Msun$ than observed, depending on the model. However, the total number of satellites depends on the mass of the host halo, generally increasing in more massive haloes. Thus, one may in principle choose a more massive one to increase the total number of satellites. This scatter driven by the variation in the host halo mass in our sample limits the constraining power of our comparison to the real MW. However, even if we had chosen haloes in a narrower mass bin, there would still be an intrinsic scatter due to different assembly histories. As shown by the error bars in Fig.~\ref{mstar_function}, which are representative of the scatter across all models, we find no significant inconsistencies with observations in the studied stellar mass range. Hence, we cannot rule out any based on stellar mass functions alone. Finally, we are not able to find LMC and SMC analogues around any of the studied haloes. This is likely because they are uncommon in isolated systems \citep{Santos.2021}.

\begin{figure*}
    \centering
    \includegraphics{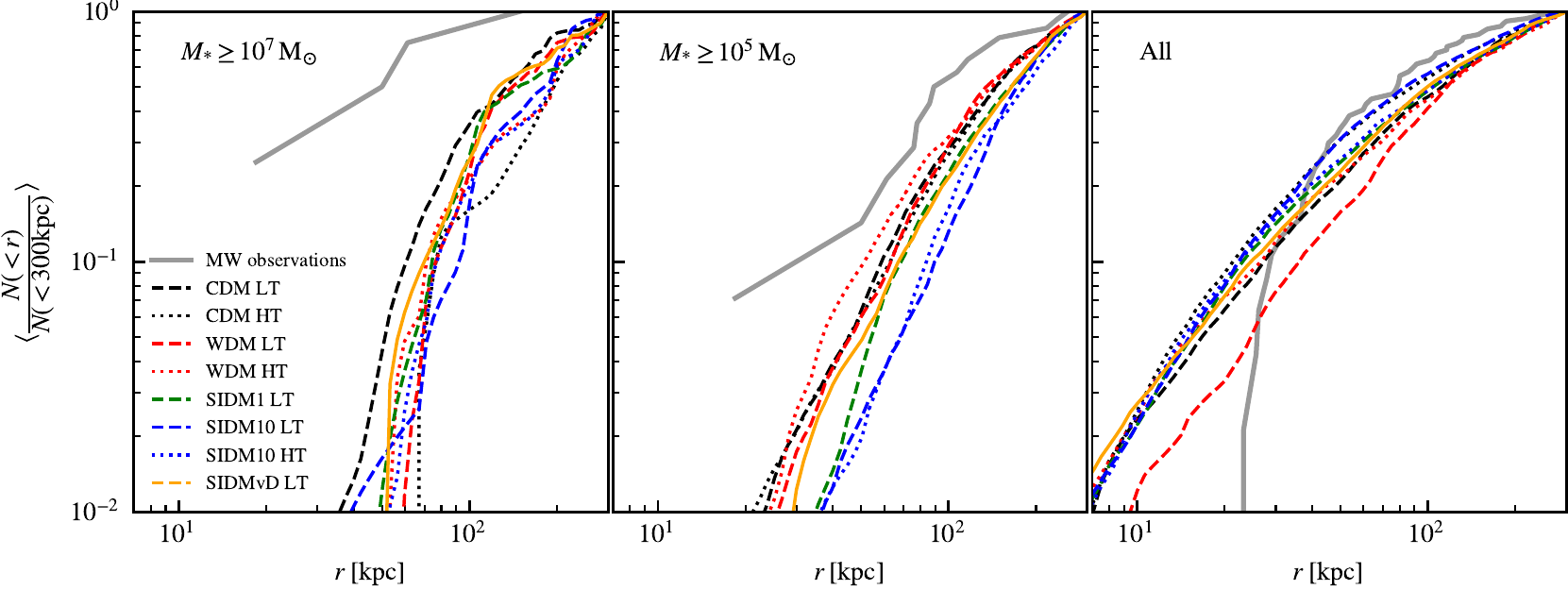}
    \caption{Radial distribution of $z=0$ satellites, averaged over all haloes in our sample over the last $300 \, \mathrm{Myr}$ of the simulation. This is shown across different mass bins, as indicated in the top left corner of each panel. The simulated sample for each bin corresponds to $M_{*}\geq 10^{7}$, all resolved, and all resolved plus orphans. The colours encode the dark matter model used, with the line styles the baryonic physics employed, as the legend indicates. The observed radial distribution of MW satellites is shown by the grey line. These values were calculated from \citet{McConnachie.2012}, assuming $R_{\odot} = 8.29 \mathrm{kpc}$.}
    \label{averaged_radial_distribution}
\end{figure*}

\subsection{Maximum circular velocities}

An alternative way to examine how strongly the satellites have been stripped is through their $V_{\rm max}$ distributions. $V_{\rm max}$ decreases faster than the stellar mass of satellites, because the latter is more concentrated than the DM, which is stripped from the outskirts. We show the $V_{\rm max}$ distributions of the resolved $z=0$ satellites in Fig.~\ref{averaged_vmax_distribution}, averaged across all haloes in a given simulation.

We do not compare to observations since the maximum circular velocity of the halo is uncertain. Other quantities accessible to observations, such as the circular velocity at the half light radius, cannot be reliably measured in these simulations given their spatial resolution. 

Similarly to the stellar mass functions, the CDM LT case represents an upper bound for all models, as it is the most resilient to tides. Although most models are similar at $\nu \geq 0.2 $, noticeable differences start to appear below that. Interestingly, the distributions for CDM HT and SIDMvD are almost exactly the same below that scale, illustrating the potentially degenerate effects that baryons and different dark matter models can have.

\subsection{Radial distributions}

Another important prediction of our simulations is the radial distribution of satellites. We explore this in Fig. ~\ref{averaged_radial_distribution}, where we show this distribution for different stellar mass bins. They were measured by integrating the satellite orbits for each halo during the last 300~Myr of the simulation, as described in Section 3. We then computed the time average and the average across the seven haloes of our sample in each simulation. The observed MW satellite radial distribution is also shown for comparison..

We see that the shape of the radial distributions depends strongly on the mass range. At high end, satellites occupy the outer regions of the halo, whereas lower mass satellites are closer to the centre. In the former, CDM LT is the most concentrated of all the models, although the distributions are noisy because of the low number of objects of these masses.

Once we start considering all the resolved satellite population (middle panel), we note that the radius enclosing half the total population -- a measure of how concentrated these systems are -- is smallest for the WDM models. This is closely followed by the CDM variations. The least concentrated satellite systems are the SIDM ones, with their concentration decreasing with increasing cross-section. In other words, even though WDM might have fewer satellites above a certain mass compared to SIDM, they are more concentrated. Increasing the density threshold for star formation leads to resolved satellite distributions that are somewhat less concentrated than their LT counterparts. The relative increase is similar for CDM, WDM and SIDM10, about $6\%$.

Correcting for orphans yields radial distribution functions that are much more centrally concentrated than the resolved satellite population. This is unsurprising, since orphans correspond to ultra-faints and populate the central regions. Generally, the inclusion of orphans makes the shapes of the radial distributions more alike across simulations and similar to that of the MW at large radii. Nonetheless, there is evidence that the MW satellite system may be more concentrated than in the simulations, perhaps due to the presence of Magellanic systems in the real MW \citep{Santos.2021}. We note that the orphan population only includes galaxies whose resolved progenitors had peak stellar masses above the baryonic particle mass of the simulations. Haloes which would have had a total stellar mass less than one particle are not counted. Thus, the orphan populations here are only a partial census of the low-mass satellites. Finally, the fraction of orphans relative to the total satellite population increases with the extent of tidal disruption of satellites.

\begin{figure*}
    \centering
    \includegraphics{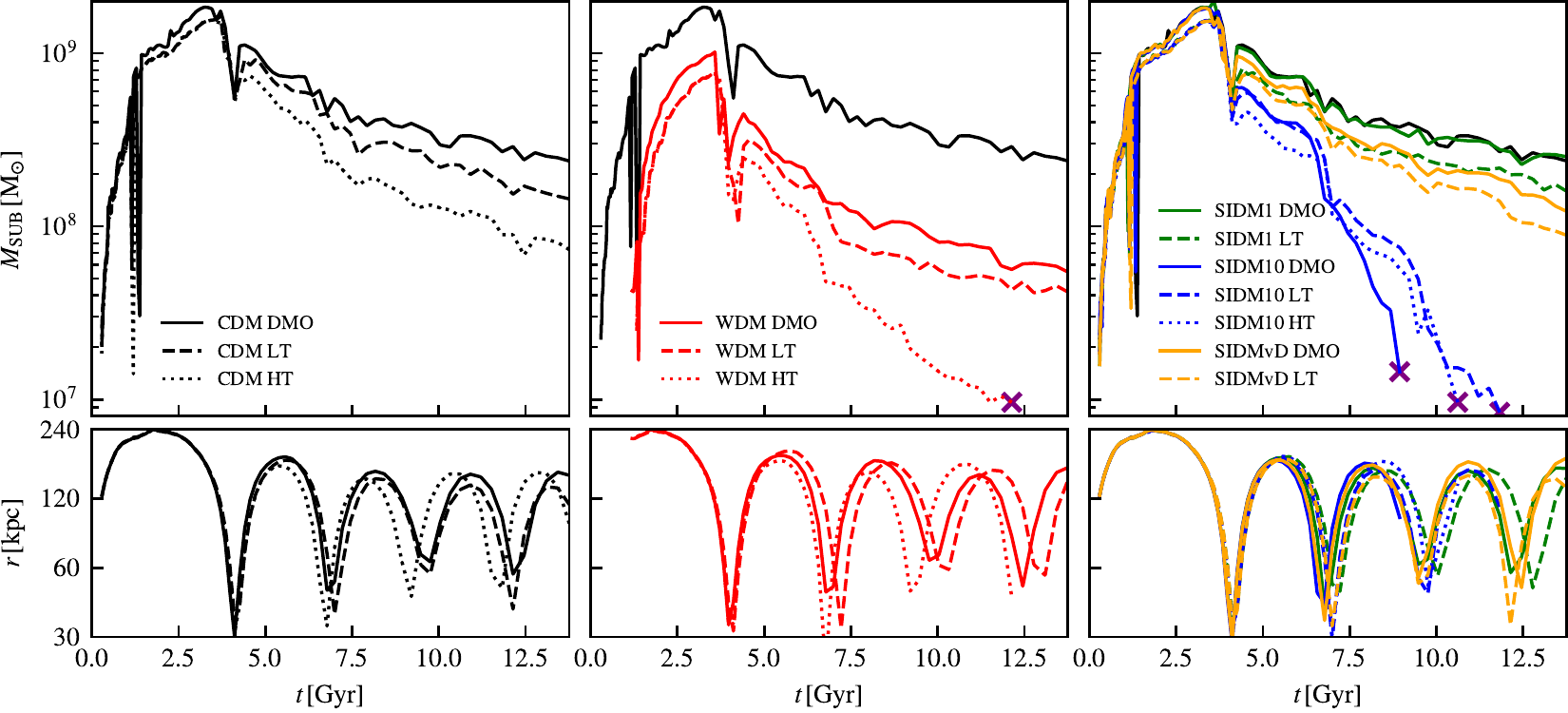}
    \caption{Time evolution of the total bound mass (top) and galactocentric distance (bottom) of a $z=0$ satellite identified in the CDM LT simulation, matched across all simulations. Each column shows the evolution in different DM models: CDM (left), WDM (centre) and SIDM (right), with the colour-coding in the latter indicating the cross-section value, as per the legend. The choice of subgrid physics is represented by the different line-styles. We show the evolution of the CDM DMO satellite in all panels to allow for an easier comparison across models. The counterparts not surviving until $z=0$ are highlighted by a purple cross at the time when they were last resolved.}
    \label{example_bound_mass_evolution}
\end{figure*}

\section{The reason behind the suppression of satellite numbers}

We have given a broad overview of how the overall population properties of the $z =0$ surviving satellies differ across models. In summary, these were variations in the number of satellites and different radial and $V_{\rm max}$ distributions. To investigate the underlying causes for these changes, we now turn to a more detailed comparison of how differences in the dark matter and baryon physics affect stripping, and thus satellite survivability.

For this, we select all satellites in the CDM LT model that are resolved at $z=0$ and identify their counterparts in the other simulations. We base our selection on this model because it has the largest surviving satellite population at $z=0$. The matching is done bijectively, as described in Section 3. In short, we identify the time at which the satellite progenitors attained their largest bound mass and cross match the 100 most bound DM particle at that time. This minimises the effects of tidal stripping and potentially diverging evolutionary paths. Moreover, this method is also able to identify counterparts that have been disrupted before $z=0$. 

We are able to find counterparts in the CDM HT and SIDM simulations for $\sim 99\%$ of the $z=0$ surviving CDM LT satellites. The number of identified counterparts in the WDM simulation is $\sim 88\%$, because the population size is smaller due to the cut-off in the power spectrum. 
\subsection{Different fates for the same satellite}

We start by considering the evolution of a single example of a satellite identified in the CDM LT simulation, whose matched counterparts retain similar orbital parameters throughout their existence. This is an important condition when comparing the evolution of a single object across simulations, as small differences in position and velocity near pericentre may lead to very different subsequent orbits and thus the tides they experience. We aim to exclude differences in stripping that are caused by changes to the orbits.

The evolution in galactocentric distance of the chosen satellite is shown in the bottom panel of Fig. \ref{example_bound_mass_evolution}. As expected, we see no differences prior to the first pericentric passage. Afterwards, we observe some minor changes to the orbital phase, but all the counterparts that survived up to $z=0$ have experienced four pericentric passages since they first entered the virial region of the halo.

Focusing on the evolution of total bound mass, there are very few differences prior to infall. At early times there are transient decreases associated with ongoing mergers, during which SUBFIND switches the subhalo it identifies as the most massive within a FoF group. We see that the peak bound mass for a fixed DM model changes between their hydrodynamical and DMO counterparts. As explained in Section 4, this is caused by the early loss of baryons and subsequent decrease in halo growth due this reduction in mass. Finally, we note the significant delay in the formation of the WDM counterparts, which lowers their peak bound mass relative to their CDM and SIDM equivalents.

The bound mass of satellites decreases continuously after infall into the virial region, with periods of intense stripping occurring near pericentre. These are often accompanied by a peak-trough-peak pattern, caused by a decrease in the tidal radius of these objects near pericentre (and thus the bound mass assigned by SUBFIND). The resulting bound mass is lowered between consecutive apocentres. A measure of how stripped these objects were by any given pericentric passage can be estimated by taking the bound mass ratio of the peaks immediately before and after a pericentric passage.

We do this for the first pericentre, which is when the orbits are most similar across simulations. The CDM DMO, LT and HT versions lost 28\%, 41\% and 54\% of their total mass, respectively. For the WDM we note a similar ordering -- but differing magnitude -- of stripping: 56\%, 63\% and 67\%. The SIDM counterparts are stripped to varying degrees depending on the cross-section. The lowest value, $1\,\mathrm{cm}^{2}\,\mathrm{g}^{-1}$, exhibits little difference to CDM, as expected since the structural changes at this mass scale are minimal (see bottom panel of Fig~\ref{density_ratios}). All of the SIDM10 versions lose a large fraction of mass, ranging from 60\% to 70\%. As expected, the stripped mass fraction in SIDMvD lies between the cases for the lowest and highest cross-sections. 

The cumulative effect of subsequent pericentres and continuous stripping leads to different subhalo masses at $z=0$. In come cases, like WDM HT or all of the SIDM10 counterparts, the mass loss causes the subhalo to be disrupted before $z=0$. For those that survive, we see a clear separation between different cross-section values and whether baryons are present or when cores form due to a high density threshold for star formation. 

To check whether the differences of $z=0$ mass between DMO and hydrodynamic counterparts is due to enhanced stripping or simply caused by a lower peak bound mass, we compute the relative loss of mass, $1 - M(z=0)/M_{\rm peak}$. For CDM, we measure $84\%$, $91\%$ and $95\%$ for the DMO, LT and HT versions, respectively. There are no differences for the WDM cases, with both the DMO and LT cases losing $94\%$ of their mass. The SIDM cases do show some differences, but less pronounced than in the CDM case; about a 3 percent increase in the mass loss rate in the hydrodynamical simulation. Note that this comparison does not attribute the increase of stripping in the hydrodynamical simulations to any single origin. Indeed, it can be caused by a combination of the presence of a massive stellar disk, the contraction of the host halo or changes in the satellite density profiles. 

\subsection{Disruption rates}

\begin{figure}
    \centering
    \includegraphics{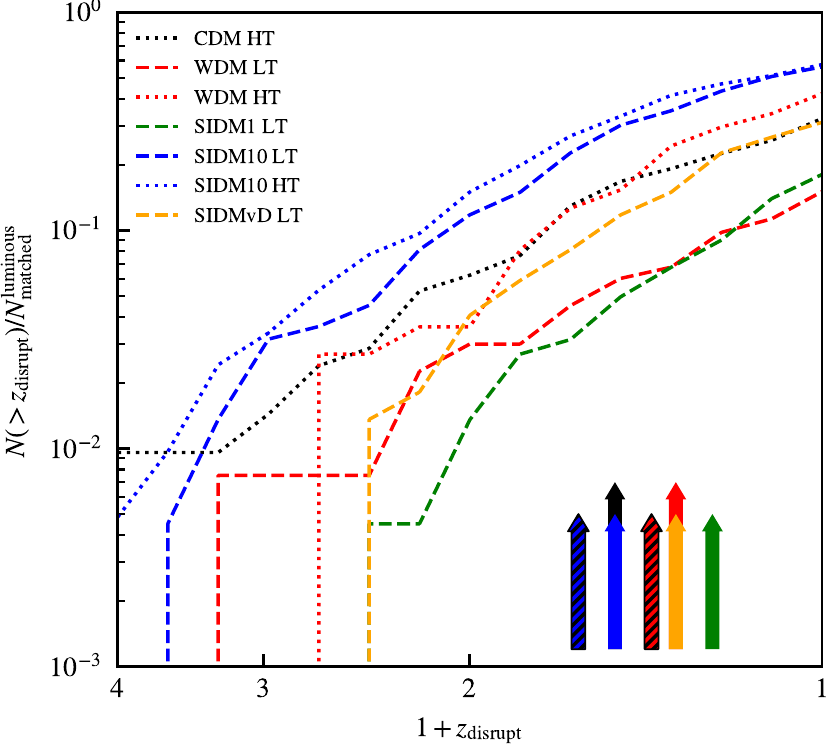}
     \caption{Cumulative distribution of the disruption redshifts for the luminous subset of matched satellites, relative to the total number of luminous objects. The colours of each line indicate the DM model as per the legend, with the dashed and dotted ones corresponding to the LT and HT versions' counterparts, respectively. The arrows on the bottom right indicate the redshift at which half of the total disrupted (luminous) population was reached. The hatched arrows correspond to the HT counterparts.}
    \label{disruption_redshift_comparison}
\end{figure}

Based on the previous example, as well as on the decrease in satellite numbers in some models even when the number of progenitors is the same, we expect many more satellites to be disrupted before $z = 0$ in the non-CDM LT counterparts. We explore this in Fig~\ref{disruption_redshift_comparison}, where we show the cumulative fraction of CDM LT counterparts in other simulations are disrupted before z = 0, as a function of the redshift when they were last resolved. This is only computed for the luminous subset of the matched populations; this does not significantly alter the SIDM and CDM HT numbers, since $\sim 96\%$ and $\sim 91\%$ of matched satellites are luminous, respectively. The difference in the SIDM case is likely caused by slight differences in the evolutionary histories, since whether or not a halo contains a single bound star particle becomes a stochastic process. In the case of CDM HT the difference stems from a combination of this and the fact that the onset of star formation will occur at later times due to the increase in the density threshold for star formation. Finally, the WDM luminous matched fractions in the LT and HT versions are $66\%$ and $55\%$, respectively. This results from the delay in the formation time of the satellite progenitors, which decreases the number of would-be satellites that cross the mass-threshold to trigger the gravitational collapse of gas (as shown by the halo occupation fraction shown in Fig~\ref{occupation_fraction}). 

Focusing on the total fraction of disrupted satellites, we observe that more than half of all satellite progenitors in the LT and HT SIDM10 simulations are disrupted before $z=0$. The lack of a significant difference between the two is likely that the gas density threshold for star formation does not alter the internal structure of these SIDM haloes significantly, unlike models where it is able to turn a cusp into a core. As the cross-section value is lowered, so is the fraction of disrupted satellites: $31\%$ and $18\%$ for the SIDMvD and SIDM1 models, respectively. About a third of all satellites in the CDM HT are disrupted before $z=0$, likely due to the structural changes caused by gas blowouts. Finally, the low threshold version of WDM loses only a small fraction of the luminous population ($\sim 13 \%$). The high threshold version, as in the other cases, exhibits a slight enhancement in disruption rates. We conclude that the decrease in the number of satellites in the WDM cosmologies, compared to SIDM, is largely due to the suppression in the number of galaxies that are able to form. 

\section{Discussion}

There are clear differences in the internal structure of dark matter haloes amongst CDM, WDM and SIDM models in DM-only simulations. However, these differences are greatly reduced by the effects of baryons. Depending on the halo mass range, choices regarding the subgrid physics may lead to comparable halo density profiles, as shown in the middle row of Fig.~\ref{density_ratios}. The similarity in the density profiles, in turn, leads to similar stripping histories. This results in degeneracies in the way in which baryon effects and the nature of the DM affect the properties of the satellite population. For example, the satellite $V_{\rm max}$ functions in SIDM with velocity-dependent cross-section and in CDM with a high density threshold for star formation are very similar (see Fig.~\ref{averaged_vmax_distribution}). 

Our analysis thus indicates that the current freedom in the modelling of star formation and feedback in simulations makes it difficult to disentangle their effects from those arising from the nature of the dark matter. Although we have considered only a subset of all the possible model variations,  our work suffices to highlight the problem and the current limitations on interpreting observational results. We thus conclude that it will be challenging to constrain the nature of dark matter based solely on the properties of $M_{*} \geq 10^{5} \Msun$ satellite systems.

A promising avenue to explore further is the population of ultra-faint satellites. As shown in the lower row of Fig. \ref{density_ratios}, the effects of baryons become increasingly less important in lower mass haloes, a direct consequence of their low baryonic content \citep{DiCintio.2014, Tollet.2016}. The internal structural differences in the halos of ultra-faint satellites, driven by the nature of dark matter, are largely preserved in the presence of baryons and affect their resilience to tidal stripping. Therefore, the properties of the ultra-faint population may retain an imprint of the nature of the dark matter. The study of these galaxies requires very high-resolution simulations that can resolve not only the formation of the faintest systems but also track their evolution as they fall into the halo and undergo tidal stripping. 

Indeed, adequate numerical resolution is essential to model tidal stripping correctly. Work based on idealised collisionless simulations suggests that cuspy dark matter haloes are resilient to tides and always leave a small bound remnant behind \citep{Errani.2021}. However, the limited resolution of cosmological simulations makes this regime difficult to follow. Furthermore, when the subhaloes are not sufficiently well resolved, the rate at which they are stripped becomes artificially high \citep{vandenBosch.2018}. Studying ultra-faint satellites and unveiling their constraining power on the nature of the dark matter will thus require very high-resolution simulations.

Our analysis relies on several assumptions. There are also limitations inherent to our simulations. Firstly, our sample selection is based solely on the virial mass of FoF groups at $z = 0$. This criterion does not account for other factors relevant to tidal stripping, such as the mass of the host's stellar component. The reference EAGLE subgrid model underpredicts the stellar mass of MW-mass haloes by a factor of $\sim 2$ \citep{Schaye.2015}. Changes to the subgrid physics alter, to some degree, the resulting stellar masses of individual haloes. However, the average stellar-to-halo-mass relation is insensitive to the models considered here. Thus, we are confident that the relative differences we observe across the average satellite populations are due to structural changes in the subhaloes rather than differences in the central galaxy properties.

Additionally, the modelling of orphans is derived from convergence studies based on the Millenium I and II simulations. These are collisionless CDM simulations, which means that the effect of baryons, self-interactions and the presence of central density cores are not considered. Their dark matter resolution is several orders of magnitude lower ($10^{9}$ and $9\times 10^{6} \Msun$, respectively) than in the simulations used in this work. This study extrapolates their findings to different DM models and higher mass resolutions. This framework will need to be extended to alternative DM models.

\section{Conclusions}

We have simulated the assembly of haloes with masses within a factor of two from the MW and their satellite systems in a cosmological setting using different dark matter models. They were run on DMO and hydrodynamical simulations via the inclusion of baryonic physics using the EAGLE subgrid model, which under certain parameter choices, can lead to the formation of baryon-driven cores. This was done with the aim of studying how these changes affect the satellite populations between pairs of matched haloes and identify systematic differences. 

Firstly, we saw significant differences at the field halo level across different simulations:
\begin{itemize}
    \item Low mass haloes in hydrodynamical simulations lose their baryons at early times, leading to delays in their formation time and, correspondingly, lower $z = 0$ virial masses.
    \item The cut-off in the power spectrum of WDM leads to a smaller number of low mass haloes forming. It also leads to a formation time delay much larger than that caused by baryons. This further suppresses the galaxy population, since the halo occupation fractions change and thus galaxy formation becomes less efficient. 
    \item The density profiles of CDM and WDM are virtually indistinguishable at high masses. They are significantly different at low masses, resulting from different concentrations that reflect their delayed formation times. 
    \item All SIDM haloes show significant differences with respect to the CDM density profiles, due to the formation of cores whose size scales with cross-section value. However, the inclusion of baryons makes the differences less apparent. At high halo masses, this is due to an interplay between self-scatterings and a contraction caused by the central galaxy. At low masses, it is caused by decreases in the overall DM density due to the delay in formation time resulting from slower growth triggered by the loss of baryons. This consequently affects the scattering rate between particles and thus the radius at which haloes are considered to have been thermalised.
\end{itemize}

All of these changes propagate to the infall properties of satellites, either via a reduction in the accreted number or structural changes that alter their subsequent evolution under the influence of tides. 
\begin{itemize}
    \item The delay in formation time of haloes, either via the loss of baryons, a cut-off in the power spectrum or both, leads to lower bound masses at infall.
    \item Structural differences across models leads to different stripping rates, which are noticeable even after just one pericentric passage.
    \item In SIDM, increasing cross-sections lead to larger cores and thus more efficient stripping. At this resolution level, the lowest cross-section value used in this study, $1 \, \mathrm{cm}^2\,\mathrm{g}^{-1}$, yields predictions with little to no difference compared to CDM.
    \item Increasing the density threshold for star formation allows the gas to accumulate in larger quantities before being blown out via supernovae feedback. This results in greater gravitational coupling to the DM, allowing it to flatten the inner DM density profile as its removed. This leads to more efficient stripping relative to their cuspy counterparts. The effect is minor in SIDM models, since haloes already have flat inner density profiles due to DM self-scattering.
\end{itemize}

The above changes lead to a suppression in the number of satellites at $z=0$ and lower $V_{\rm max}$ values for those surviving. In SIDM, this is solely caused by the enhanced stripping as a consequence of their flat inner DM density profiles. The lack of satellites in WDM is almost entirely attributable to less haloes (and galaxies) forming in the first place. Models in which gas blowouts are able to flatten the density profiles of haloes also show a suppression in satellite numbers, even in CDM and WDM. In some cases, they lead to entirely degenerate satellite system properties, such as the stellar mass distribution in CDM with baryon-driven cores and velocity-dependent SIDM.

In summary, despite differences amongst dark matter models in DM-only simulations, the presence of baryons can erase the differences arising due to the nature of the dark matter. Our analysis demonstrates that the study of the satellite population in the mass range of $M_{*} \geq 10^{5}\, \Msun$ is unlikely set informative constraints on the nature of the DM. The lack of constraining power of massive satellites, however, does not rule out the possibility that less massive systems, particularly ultra-faint dwarfs, could be sensitive to the properties of the DM. Understanding and quantifying these constraints  will require the development of dedicated, extremely high-resolution cosmological simulations, an endeavour worth pursuing.

\section*{Acknowledgements}
ABL acknowledges support by the European Research Council (ERC) under the European Union's Horizon 2020 research and innovation programme (GA 101026328) and UNIMIB's Fondo di Ateneo Quota Competitiva (project 2020-CONT-0139). CSF, SMC and VJFM acknowledge support by the European Research Council (ERC)
through Advanced Investigator grant DMIDAS (GA 786910) and Consolidated Grant ST/T000244/1. This work used the DiRAC@Durham facility managed by the Institute for Computational Cosmology on behalf of the STFC DiRAC HPC Facility (www.dirac.ac.uk). The equipment was funded by BEIS capital funding via STFC capital grants ST/K00042X/1, ST/P002293/1, ST/R002371/1 and ST/S002502/1, Durham University and STFC operations grant ST/R000832/1. DiRAC is part of the National e-Infrastructure.
%%%%%%%%%%%%%%%%%%%%%%%%%%%%%%%%%%%%%%%%%%%%%%%%%%
\section*{Data Availability}

The data used in this study can be made available upon reasonable request to the corresponding author.

%%%%%%%%%%%%%%%%%%%% REFERENCES %%%%%%%%%%%%%%%%%%

\bibliographystyle{mnras}
\bibliography{references}
%%%%%%%%%%%%%%%%%%%%%%%%%%%%%%%%%%%%%%%%%%%%%%%%%%

% Don't change these lines
\bsp	% typesetting comment
\label{lastpage}
\end{document}